\documentclass[prd,aps,twocolumn]{revtex4}
\usepackage{amssymb,amsmath,amsfonts,amsbsy,epsfig}

\newcommand{\be}{\begin{equation}}
\newcommand{\ee}{\end{equation}}
\newcommand{\bea}{\begin{eqnarray}}
\newcommand{\eea}{\end{eqnarray}}
\newcommand{\nn}{\nonumber}

\def\k{{\bf k}}

\def\x{{\bf x}}

\def\z{{\bf z}}

\begin{document}

\title{Axion excursions of the landscape during inflation}

\author{
Gonzalo A. Palma and Walter Riquelme
}

\affiliation{
Grupo de Cosmolog\'ia y Astrof\'isica Te\'orica, Departamento de F\'{i}sica, FCFM, \mbox{Universidad de Chile}, Blanco Encalada 2008, Santiago, Chile.}

\begin{abstract} 

Because of their quantum fluctuations, axion fields had a chance to experience field excursions traversing many minima of their potentials during inflation. We study this situation by analyzing the dynamics of an axion-spectator field $\psi$, present during inflation, with a periodic potential given by $v(\psi) = \Lambda^4 [1 - \cos (\psi / f)]$. By assuming that the vacuum expectation value of the field is stabilized at one of its minima, say $\psi = 0$, we compute every $n$-point correlation function of $\psi$ up to first order in $\Lambda^4$ using the in-in formalism. This computation allows us to identify the distribution function describing the probability of measuring $\psi$ at a particular amplitude during inflation. Because $\psi$ is able to tunnel between the barriers of the potential, we find that the probability distribution function consists of a non-Gaussian multimodal distribution such that the probability of measuring $\psi$ at a minimum of $v(\psi)$ different from $\psi=0$ increases with time. As a result, at the end of inflation, different patches of the Universe are characterized by different values of the axion field amplitude, leading to important cosmological phenomenology: (a) Isocurvature fluctuations induced by the axion at the end of inflation could be highly non-Gaussian. (b) If the axion defines the strength of standard model couplings, then one is led to a concrete realization of the multiverse. (c) If the axion corresponds to dark matter, one is led to the possibility that, within our observable Universe, dark matter started with a nontrivial initial condition, implying novel signatures for future surveys.

\end{abstract}

\maketitle 

\section{Introduction}

How did cosmic inflation~\cite{Guth:1980zm, Linde:1981mu, Albrecht:1982wi, Starobinsky:1980te, Mukhanov:1981xt} influence the structure of the standard model of particle physics? The string-landscape picture~\cite{Susskind:2003kw}---in which the standard model (SM) is just one of many possible vacua---provides us with a useful framework to address this compelling question. In it, the inflationary history plays a crucial role in selecting the properties of our Universe, as we observe it~\cite{Linde:2015edk}. 

This selection could have been classical if the inflationary trajectory followed by the scalar fields respected classical equations of motion. In this case, the SM vacuum would have been determined by certain initial conditions specifying the starting point of the inflationary attractor trajectory. In these situations the probability distribution functions (PDFs) describing any scalar degrees of freedom remained nearly Gaussian, and the analysis of their role on cosmological observables is limited to the computation of two- and three-point correlation functions.

Another more intriguing possibility is that, during inflation, tunneling across barriers separating different classical trajectories was relevant. In this case the vacuum selection was intrinsically quantum mechanical, and the quantum to classical transition during inflation~\cite{Polarski:1995jg, Lesgourgues:1996jc, Kiefer:2008ku, Burgess:2014eoa} would have played a decisive role in determining the properties of the standard model. Moreover, the PDFs of extra scalar fields would be highly non-Gaussian, and an analysis based on the computation of two- and three-point correlation functions is found to be insufficient to understand their role at the end of inflation.

In this paper, we study the role of quantum fluctuations in determining the final state of inflation and so the properties of particle physics within our observable Universe. To do this, we analyze a very simple inflationary setup containing an axion spectator field~\cite{Peccei:1977hh} with a sinusoidal periodic potential. This setup is simple enough to allow us perturbative computations leading to nontrivial results that are representative of more complicated systems, containing other classes of fields and/or potentials. We will show that the quantum fluctuations of axions are able to traverse many local minima of their periodic potentials, leading to a non-Gaussian multimodal PDF. In consequence, the final value at which they stabilize remains indeterminate during inflation, leading to a universe filled with patches characterized by different vacuum expectation values (VEVs) of the axion. Our results should lead to the derivation of new constraints on axions (and spectator scalar fields in general) connecting current cosmological and/or astrophysical constraints with primordial cosmology.

\section{Axions in inflation}

Axions may have an important role in connecting particle physics with cosmology~\cite{Baumann:2014nda}. Axions are natural candidates for the inflaton~\cite{Freese:1990rb, McAllister:2008hb}, but this requires them to have super-Planckian decay constants, unless some alignment effect underlies their potentials~\cite{Kim:2004rp}. In addition, thanks to axions, inflation may play a role in the existence of hierarchies in the SM~\cite{Graham:2015cka}. 

In this paper, we will not bother about the specific nature of the inflaton. We will consider a model in which there is an axion field $\psi$ that remains decoupled (at tree level) from the inflaton field $\phi$, in charge of driving inflation. This class of models has been studied in the past~\cite{Seckel:1985tj, Linde:1987bx, Turner:1990uz}, with an emphasis on the production of Gaussian isocurvature perturbations. The action describing this system is given by
\be
S =  S_{\rm EH} [g_{\mu \nu}]   +  S_{\rm infl} [g_{\mu \nu}, \phi]  +  S_{\psi} [g_{\mu \nu} , \psi]  , \label{general_action}
\ee
where $S_{\rm EH}  [g_{\mu \nu}] $ corresponds to the Einstein-Hilbert action describing the dynamics of the metric $g_{\mu \nu}$ and $S_{\rm infl} [g_{\mu \nu}, \phi]$ and  $S_{\psi} [g_{\mu \nu} , \psi]$ are the respective actions for $\phi$ and $\psi$, both of them minimally coupled to gravity. The action $S_{\rm infl} [g_{\mu \nu}, \phi]$ includes a potential $V_0 (\phi)$ that produces inflation. The inflationary background is well described by a Friedman-Robertson-Walker metric of the form $ds^2 = - dt^2 + a^2 (t) d \x^2$, where $a(t)$ is the scale factor. If $V_0 (\phi)$ is flat enough, the scalar field $\phi$ rolls down the potential slowly, and the background is well approximated to a de Sitter geometry, with deviations of the order of $\epsilon \equiv - \dot H / H^2$, where $H = \dot a / a$ is the Hubble expansion rate. The latest cosmic microwave background (CMB) observations tell us that $H_* < 8.8 \times 10^{13}$GeV~\cite{Ade:2015lrj}, where $H_*$ is the Hubble expansion rate at the time when fluctuations of wave number $k_* = 0.05$Mpc$^{-1}$ crossed the horizon.

The axion $\psi$ has a potential $v(\psi)$ that we take to be
\be
v(\psi)   \equiv   \Lambda^4 \left[1 - \cos \Big(\frac{\psi}{f} \Big) \right] , \label{pot-axion}
\ee
where $f$ is the axion decay constant. This potential is the result of nonperturbative effects due to the coupling between the axion and gauge fields. It is convenient to write the action for $\psi$ in conformal coordinates and in terms of a canonically normalized field $u = a \psi$. We may define conformal time $\tau$ through $d\tau = dt / a$. Then, the action $S_{\psi}$ of Eq.~(\ref{general_action}) takes the form
\be
S_{\psi} = \int \!\! d^3x \, d\tau \left[   \frac{1}{2} ( u ')^2 + \frac{1}{2 } (\nabla u)^2 - a^4 v(u/a)   \right] .
\ee
In the de Sitter limit $\epsilon \to 0$ one has $H=$const. In this case one has $a (\tau) = - 1 / H \tau$. The conformal time covers the range $-\infty < \tau < 0$, and the limit $\tau \to 0^{-}$ corresponds to $t \to + \infty$.

\section{The in-in formalism}

In what follows, we focus on the computation of $n$-point correlation functions of $\psi$ to first order in $\Lambda^4$. We shall use the shorthand notations $\int_x = \int d^3x$ when integration takes place in coordinate space and $\int_k = (2\pi)^{-3} \int d^3 k$ when integration takes place in momentum space. In addition, we shall use the following convention to relate fields in both spaces:
\be
u (\x,\tau) = \int_k\, \hat u (\k, \tau) e^{i \k \cdot \x} .
\ee
The computation of $n$-point correlation functions in coordinate space, using the \emph{in-in} formalism, takes the form
\be
\langle u (\x_1, \tau)  \ldots u(\x_n , \tau) \rangle = \langle 0 | U^\dag  u_I (\x_1, \tau)  \ldots u_I(\x_n , \tau)  U | 0 \rangle,  \label{n-point-gen}
\ee
where $U = U(\tau)$ is the propagator and $u_I (\x, \tau)$ is the interaction picture field. This field has a momentum-space representation given by 
\be
\hat u_I (\k, \tau) \equiv a_\k u^I_k(\tau)  + a_{-\k}^{\dag} u^{I *}_k(\tau) \label{C-and-A-expansion}
\ee
where $a_{\k}^{\dag}$ and $a_{\k}$ are creation and annihilation operators satisfying the standard commutation relation $\big[ a_{\k} , a_{\k'}^{\dag} \big] = (2 \pi)^3 \delta^{(3)} (\k - \k')$. On the other hand, the mode functions $u_{k}(\tau)$ satisfy the equations of motion of a free field (that is, with $\Lambda = 0$). These are given by mode solutions respecting Bunch-Davies initial conditions:
\be
u^I_{k} ( \tau ) = \frac{1}{\sqrt{2 k}} \left( 1 - \frac{i}{k \tau} \right) e^{-i k \tau} . \label{u-k-BD}
\ee 
The vacuum state $| 0 \rangle$ is normalized $\langle 0 | 0 \rangle = 1$, and it is defined to satisfy $a_{\k} | 0 \rangle = 0$. The propagator may be written in terms of $u_I (\x, \tau)$ in the following way:
\be
U (\tau) = \mathcal T \exp \left\{  - i \int^{\tau}_{- \infty^{+}} \!\!\!\!\!\!\!\!\! d\tau' H_{I} (\tau') \right\} ,
\ee
where $H_{I} (\tau)$ is the Hamiltonian in the interaction picture. In the previous expression, $\mathcal T$ stands for the standard time ordering symbol, and $\infty^{+} = (1 + i \epsilon) \infty$ is the prescription isolating the in vacuum in the infinite past. In the particular case of (\ref{pot-axion}), the Hamiltonian $H_{I} (\tau)$ takes the form
\be
H_{I} (\tau) = \frac{\Lambda^4}{H^4 \tau^4} \int_z \left[ 1 - \cos  \Big( \frac{H \tau}{f} u_I (\z , \tau ) \Big)  \right] .
\ee
We may expand $U$ to first order in $\Lambda^4$ to compute $n$-point correlation functions. Then, in momentum space Eq.~(\ref{n-point-gen}) takes the form
\bea
\langle \hat u (\k_1, \tau)  \ldots \hat u (\k_n, \tau) \rangle  =  \langle 0 |  \hat u_I (\k_1, \tau)  \ldots \hat u_I (\k_n, \tau) | 0 \rangle \quad  \nn  \\
 - i \int^{\tau}_{- \infty} \!\!\!\!\!\!  d\tau'  \langle 0 | \left[  H_{I} (\tau')  , \hat u_I (\k_1, \tau)  \ldots \hat u_I (\k_n, \tau)  \right] | 0 \rangle .  \quad \quad  \label{n-point-first}
\eea
We have dropped the prescription involving $\infty^{+}$ which is irrelevant for computations up to first order in the interaction. 

It is important to notice that the Bunch-Davies initial condition of Eq. (\ref{u-k-BD}) does not share the periodicity of the potential. For this initial condition to be valid, we have to assume that the wall domain number $N$ associated to the axion field is much larger than 1.

Before analyzing the entire theory in detail, let us briefly examine the trivial case in which $\Lambda = 0$. In this case the theory corresponds to a massless field in a de Sitter space-time, and the distribution function describing the probability of measuring a specific amplitude is Gaussian. The variance of such a distribution is given by the two-point correlation function $\sigma_0^2 (\tau) \equiv \langle \psi^2 (\x , \tau) \rangle$ (with $\x$ evaluated at any desired value), which in terms of $u_\k$ has the form
\be
\sigma_0^2  = H^2 \tau^2 \int_k u_{k} (\tau) u_{k}^* (\tau) . \label{sigma-0}
\ee
Notice that $\sigma_0^2$ is time independent, which is possible to verify by absorbing the combination $- k \tau$ into a single integration variable. This time independence comes from the fact that we are dealing with fluctuations in a de Sitter space-time. On the other hand, $\sigma_0^2$ is formally infinite on account of (\ref{u-k-BD}). This indeterminacy may be eliminated by introducing both infrared and ultraviolet cutoffs. These details will turn out to be irrelevant in our analysis.

\section{Computing $n$-point functions}

\begin{figure}[t!]
\includegraphics[scale=0.45]{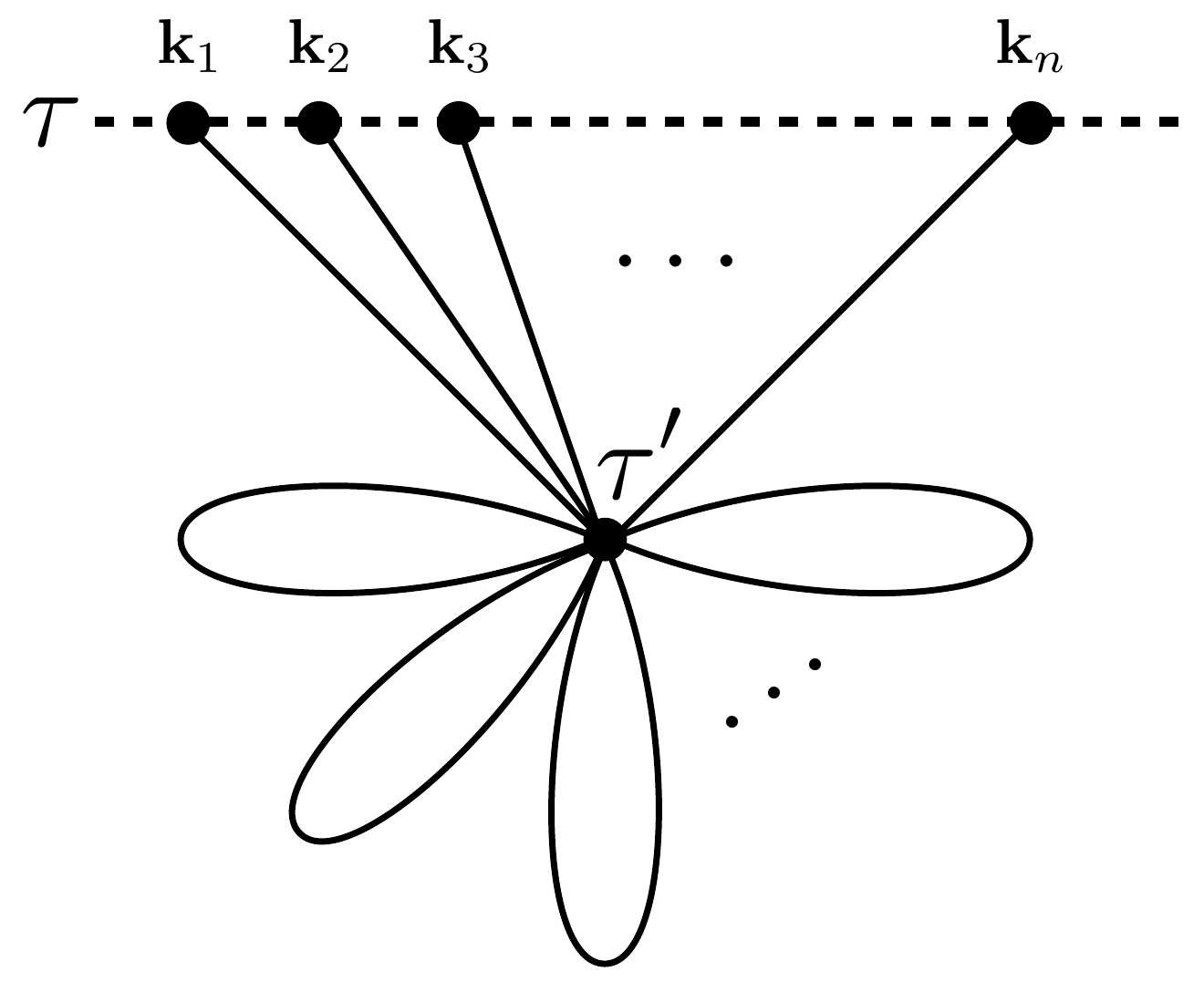}
\caption{A typical connected diagram of the order of $\Lambda^4$. All the $n$-external legs are attached to one of the available $2m$ legs of the vertex. The remaining legs from the vertex become loops.}
\label{fig:FIG1}
\end{figure}

We now proceed to compute the second line in Eq.~(\ref{n-point-first}), which corresponds to the non-Gaussian contribution to the $n$-point correlation function. Notice that this computation will take into account the leading-order contribution in terms of $\Lambda^4$ which involves the whole function $1- \cos(\psi /f)$. In this computation, we will encounter the following function:
\be
\Delta( \tau' , \tau , k) \equiv   u^I_{k} (\tau') u^{I*}_{k} (\tau) ,
\ee
which may be thought of as a propagator in momentum space connecting vertices characterized by $\tau$ and $\tau'$. Now, to compute~(\ref{n-point-first}), we first expand the cosine function appearing in the potential. This leads to a Hamiltonian of the form
\be
H_I (\tau)  = - \frac{\Lambda^4}{H^4 \tau^4}  \sum_{m=1}^{\infty} \frac{(-1)^m}{(2m)!} \int_z \left( \frac{H \tau}{f} u_I (\z, \tau)\right)^{2m}  .
\ee
By plugging this expression back into Eq.~(\ref{n-point-first}), we find that the contribution proportional to $\Lambda^4$ is given by
\bea
\langle \hat u (\k_1, \tau)  \ldots \hat u (\k_n, \tau) \rangle_{\Lambda^4}  = i \frac{\Lambda^4}{f^4} \sum_{m=1}^{\infty} \frac{(-1)^m}{(2m)!} \qquad \qquad \qquad \nn \\ 
\qquad \qquad \int^{\tau}_{- \infty} \!\!\!\!\!\!  d\tau'  \left( \frac{H \tau'}{f}\right)^{2m-4} F(\tau' , \tau, \k_1, \ldots \k_n) , \quad \quad  \label{n-point-intermediate-1}
\eea
where we have defined the function $F(\tau' , \tau, \k_1, \ldots \k_n)$ as
\be
F \equiv \int_z  \langle 0 | \left[ \left( u_I (\z, \tau')\right)^{2m}  ,  \hat u_I (\k_1, \tau)  \ldots \hat u_I (\k_n, \tau)  \right] | 0 \rangle . \label{F-def}
\ee
Because of invariance under spatial translations, this function is proportional to a $\delta^{(3)}(\sum_j \k_j)$. In addition, it is nonvanishing only for even values of $n$. The function $F$ may be represented as a sum of diagrams with a vertex located in $\tau'$ with $2m$ legs, some of these connected to the $n$-external legs labeled by the momenta $\k_i$ at a time $\tau$ (see Fig.~\ref{fig:FIG1}). It will be enough for us to consider the fully connected diagrams, in which every external leg is attached to the vertex. Let us refer to these contributions as $F_c$. It is clear that $F_c \neq 0$ only if $2m \geq n$. The number of ways in which the $n$ external legs may be connected to the $2m$ vertex legs is $(2m)!/(2m-n)!$. In addition, $F_c$ will have $m - n/2$ loops resulting from vertex legs that are not attached to any of the external legs. The number of different ways in which such legs can be connected into loops is given by $(2m - n)!/[2^{m-n/2} (m-n/2)!]$. 
All of this tells us that $F_c$ will have a combinatorial factor $(2m)! / [2^{m-n/2} (m-n/2)!]$ due to the different possible ways of contracting the various interaction fields $ u_I$ present in Eq.~(\ref{F-def}). More precisely, we find
\bea
F_c  = &&  \!\!\!\! - i \, (2 \pi)^3 \delta^{(3)}
\Big( \sum_j \k_j \Big) \frac{ (2m)!  }{2^{m-n/2} (m-n/2)!} \nn \\ 
&& \,\, \left[ \int_k \! \Delta( \tau' , \tau ' , k) \right]^{m-n/2}  \!\! G_c  (\tau' , \tau, k_1, \ldots k_n), \qquad \label{F-c-partial}
\eea
where we have defined $G_c (\tau' , \tau, k_1, \ldots k_n)$ as 
\bea
G_c (\tau' , \tau, k_1, \ldots k_n) = && \!\!\!\! i \sum_{l=1}^{n}   \Delta( \tau , \tau ' , k_1) \ldots \Delta( \tau , \tau ' , k_{l-1})   \nn \\
&& \!\!\!\! \left[ \Delta( \tau' , \tau  , k_l) - \Delta( \tau , \tau ' , k_l)  \right]�\nn \\
&&  \!\!\!\! \Delta( \tau' , \tau , k_{l+1})  \ldots \Delta( \tau' , \tau , k_{n}) . 
\eea
Notice that $\int_k \Delta( \tau' , \tau ' , k) $ in Eq.~(\ref{F-c-partial}) represents a closed loop. We may now plug $F_c$ back into Eq.~(\ref{n-point-intermediate-1}) to obtain an expression for the connected contribution to the $n$-point correlation function:
\bea
 \langle u (\k_1, \tau)  \ldots u(\k_n , \tau) \rangle_c  =   (-1)^{n/2} \frac{\Lambda^4}{H^4 }
(2 \pi)^3 \delta^{(3)} \Big(\sum_j \k_j \Big)   \nn \\ 
   \sum_{m=n/2}^{\infty} \frac{1}{(m-n/2) !} \left[ - \frac{1}{2}\left( \frac{H \tau'}{f} \right)^{2} \! \int_k \Delta( \tau' , \tau ' , k) \right]^{m-n/2} \nn \\
\int^{\tau}_{- \infty} \!\!\!\!\!\!  d\tau' \frac{1}{(\tau')^4} \! \left( \frac{H \tau'}{f} \right)^{n}  \!\! G_c (\tau' , \tau, k_1, \ldots k_n) .  \,\, \qquad  \label{n-point-intermediate-2}
\eea
Now, one should notice that the $m$ sum, which comes from the expansion of the cosine, involves only closed loop contributions. If one defines $m' = m-n/2$, this sum becomes
\be
\sum_{m'} \frac{1}{m' !} \left[ - \frac{1}{2}\left( \frac{H \tau'}{f} \right)^{2} \! \int_k \Delta( \tau' , \tau ' , k) \right]^{m'} = e^{ - \frac{\sigma_0^2}{2 f^2} } ,
\ee
where we have used (\ref{sigma-0}) to identify $\sigma_0$. Thus, the resummation of all the loop diagrams leads to the following $n$-point correlation function
\bea
 \langle u (\k_1, \tau)  \ldots u(\k_n , \tau) \rangle_c  =   (-1)^{n/2} 
(2 \pi)^3 \delta^{(3)} \Big(\sum_j \k_j \Big)   \nn \\ 
\frac{\Lambda^4}{H^4 }   e^{ - \frac{\sigma_0^2}{2 f^2} } \!\! \int^{\tau}_{- \infty} \!\!\!\!\!\!  d\tau' \frac{1}{(\tau')^4} \! \left( \frac{H \tau'}{f} \right)^{n}  \!\! G_c (\tau' , \tau, k_1, \ldots k_n) .  \,\,  \label{n-point-second}
\eea
What remains is to solve the $\tau'$ integral. This integral is hard to solve in general, but we may find a useful expression valid in the limit where the $\k_i$ momenta are soft. More precisely, the integration may be divided into two regions $\tau' \in (-\infty , \tau_0)$ and $\tau' \in [\tau_0 ,\tau ]$, with $\tau_0$ chosen in such a way that $k_j |\tau_0| \ll 1$. The first contribution remains finite due to the oscillatory nature of $G_c$ in the asymptotic limit $\tau' \to - \infty$. On the other hand, the second contribution has a piece that diverges as $\tau \to 0$ given by
\be
\int^{\tau}_{\tau_0} \!\!\!  d\tau' \tau^n \left(\tau' \right)^{n-4}  G_c  \to \frac{1}{3} \frac{ k_1^3 + \cdots + k_n^3 }{ 2^{n-1}  k_1^3 \ldots k_n^3 }  \ln \left( \frac{\tau_0}{\tau} \right).
\ee
As long as the momenta satisfy $k_j |\tau_0| \ll 1$, this expression dominates the $\tau'$ integral in the limit $| \tau | \ll | \tau_0 |$.
This allows us to finally arrive to an expression for the $n$-point correlation functions:
\bea
\langle u (\k_1, \tau)  \ldots u(\k_n , \tau) \rangle_c  =  (-1)^{n/2}  (2\pi)^3 \delta^{(3)}
\Big(\sum_j \k_j \Big) \nn \\ 
 \frac{\Lambda^4}{H^4 }  e^{ - \frac{\sigma_0^2}{2 f^2} }  \left( \frac{H}{f \tau} \right)^{n} \frac{1}{3} \frac{ k_1^3 + \cdots + k_n^3 }{ 2^{n-1}  k_1^3 \ldots k_n^3 }  \ln \left( \frac{\tau_0}{\tau} \right). \,\, \label{n-point-third}
\eea
This result gives us the connected contribution to the $n$-point correlation functions in momentum space for superhorizon fluctuations (with $|\tau_0| k_j \ll 1$). It captures the effects of the cosine potential to first order in $\Lambda^4$ but to all orders in $1/f$.

\section{Probability distribution function in the long-wavelength limit}

Let us recall that cosmological observables connected to inflation, via hot big-bang era initial conditions, are determined by superhorizon fluctuations. Fortunately, we can use Eq.~(\ref{n-point-third}) to compute $n$-point correlation functions in coordinate space as long as we focus on long-wavelength contributions.  To this end, we may decompose $\psi$ into short- and long-wavelength contributions as $\psi = \psi_{\rm S} + \psi_{\rm L}$, where $\psi_{\rm L}$ contains the contributions from momenta satisfying $k |\tau_0| \ll 1$. Then, given that (\ref{n-point-third}) is valid only for soft momenta, we may use it to compute $n$-point functions $\langle \psi_{\rm L} (\x_1) \ldots \psi_{\rm L} (\x_n) \rangle_c$. In the particular case where the proper distances $L_{ij} (\tau) = |\x_i - \x_j|/H |\tau|$ are of the order of $H^{-1}$ or smaller, at any time $\tau > \tau_0$, it makes no difference to evaluate all the coordinates $\x_j$ at a common value, say $x_j = 0$. Thus, we may compute $\langle \psi_L^n \rangle_c \equiv \langle \psi_{\rm L} (\x_1) \ldots \psi_{\rm L} (\x_n) \rangle_c \big|_{\x_j \to 0}$, which, after using (\ref{n-point-third}), is found to be
\be
\langle \psi_{\rm L}^n \rangle_c = (-1)^{n/2} n \frac{A^2}{\sigma_{\rm L}^2}   e^{ - \frac{\sigma_{\rm L}^2}{2 f^2} }   \left( \frac{\sigma_{\rm L}^2}{f} \right)^{n}  \label{n-point-long}
\ee
(recall that we are assuming even values of $n$), where we have defined
\be
A^2 \equiv  \frac{ \Lambda^4}{3 H^2}   \ln \left( \frac{\tau_0}{\tau} \right) e^{ - \frac{\sigma_{\rm S}^2}{2 f^2} }  .
\ee
In the previous expressions, we have introduced $\sigma_{\rm S}^2$ and $\sigma_{\rm L}^2$ as the short- and long-wavelength contributions to the variance, in such a way that $\sigma_0^2 = \sigma_{\rm S}^2 + \sigma_{\rm L}^2$. In particular,
\be
\sigma_L^2 = H^2 \tau^2 \int_{k_L} u^I_{k} (\tau) u^{I*}_{k} (\tau) ,
\ee
where the label $k_L$ tells us that we are integrating in a range such that $k \ll |\tau_0|^{-1} $. Now, we may wonder about what class of PDF gives $n$-point correlation functions such as those of Eq.~(\ref{n-point-long}). To be precise, there must exist a probability distribution function $\rho(\psi)$ such that 
\be
\langle \psi_{\rm L}^n \rangle=\int d\psi\  \psi^{n}\rho(\psi) . \label{comp-psi^n-pdf}
\ee
Notice that $\langle \psi_{\rm L}^n \rangle_c$ shown in Eq.~(\ref{n-point-long}) gives the connected part of $\langle \psi_{\rm L}^n \rangle$. This is the contribution coming from the right-hand side of (\ref{comp-psi^n-pdf}) that brings the higher number of powers of $1/f$. This is because each external leg in the diagrammatic expansion of $\langle \psi (\x_1) \ldots \psi (\x_n) \rangle$ performed in the previous section, carries a factor $1/f$ when it is connected to a vertex of the order of $\Lambda^4$. To find $\rho (\psi)$, we may proceed by adopting the following trick. We first rewrite Eq.~(\ref{n-point-long}) in the following way:
\bea
\langle \psi_{\rm L}^n \rangle_c  \! &=& \!\!\!  \left(1 + 2f^2  \frac{\partial}{\partial \sigma_{\rm L}^2} \right)  \!\!\left[(-1)^{n/2} \frac{A^2}{2f^2}e^{ - \frac{\sigma_{\rm L}^2}{2 f^2} }  \!\! \left( \frac{\sigma_{\rm L}^2}{f} \right)^{\!\! n} \right] \! . \qquad \label{psi-xi}
\eea
Now, instead of looking for a PDF that gives us back (\ref{n-point-long}), we may look for a distribution $\xi(\psi)$ such that it gives us back the expression inside the square brackets:
\be
\int d\psi \psi^n \xi(\psi) \Big|_{c}=(-1)^{n/2} \frac{A^2}{2f^2}   e^{ - \frac{\sigma_{\rm L}^2}{2 f^2} } \left( \frac{\sigma_{\rm L}^2}{f} \right)^{n} ,
\label{integration_xi}
\ee
where the subscript $c$ denotes that we are keeping only the connected part after the integration is performed. It should be clear from (\ref{comp-psi^n-pdf})--(\ref{integration_xi}) that the relation between $\rho$ and $\xi$ is given by
\be
\rho(\psi) \Big|_{c} =\xi(\psi) \Big|_{c} +2f^2 \frac{\partial}{\partial \sigma_{\rm L}^2}\xi(\psi) \Big|_{c} . \label{rho-xi}
\ee
Next, it is not difficult to see that the part of the distribution $\xi(\psi)$ that gives the desired connected contribution must be proportional to the combination $\exp (- \psi^2 / 2\sigma_{\rm L}^2 ) \cos (\psi /f)$. This implies that
\be
\xi(\psi) \Big|_{c}=\frac{1}{\sqrt{2\pi \sigma_{\rm L}^2}}e^{-\frac{\psi^2}{2\sigma_{\rm L}^2}} \frac{A^2}{2f^2}\cos\left(\frac{\psi}{f} \right)  . \label{xi(psi)}
\ee
Now, putting together Eqs.~(\ref{rho-xi}) and (\ref{xi(psi)}), and taking into account the contributions from the omitted disconnected diagrams (which, after all, come from the Gaussian free part of the theory), we finally find that the desired PDF is exactly given by
\be
\rho (\psi) = \frac{e^{- \frac{\psi^2}{2 \sigma_{\rm L}^2}}}{\sqrt{2\pi} \sigma_{\rm L}}  \left[ 1 - A^2 \left( \frac{ \sigma_{\rm L}^2 - \psi^2 - \sigma_{\rm L}^4 / f^2 }{2  \sigma_{\rm L}^4} \right)   \cos\left( \frac{\psi}{f} \right) \right] . \label{distribution-1}
\ee
This is one of our main results. It describes the PDF of measuring the amplitude of $\psi$ at a given value. Equation (\ref{distribution-1}) is valid as long as the second term inside the square brackets remains small compared to unity. Besides this limitation, the result tells us that the probability is larger at those values that minimize the cosine potential, as it should.

\section{Discussion}

Let us do some guesswork in order to generalize Eq.~(\ref{distribution-1}) to the case in which the second term inside the square brackets is allowed to be large. If these terms come from exponentials, we are immediately led to the following possible resummation:
\be
\rho (\psi) = \frac{1}{\mathcal N} \frac{e^{- \frac{\psi^2}{2 \sigma^2(\psi) }} }{  \sqrt{2\pi} \sigma(\psi)} \exp\left[  \frac{A^2}{2 f^2} \cos (\psi / f) \right] ,  \label{distribution-2}
\ee
where $\mathcal N$ is a normalization constant that depends on $\sigma_L$, $f$, and $A$ and $\sigma(\psi)$ is a function given by
\be
\sigma(\psi) \equiv \sigma_{\rm L} \exp\left[  \frac{A^2}{2\sigma_{\rm L}^2} \cos (\psi / f) \right] .
\ee
There are good reasons to think that Eq.~(\ref{distribution-2}) is the correct PDF valid for all values of the parameter $A$, keeping in mind that we are interested in the description of long-wavelength modes. If this is the case, then (\ref{distribution-2}) would be the result of taking into account those terms in Eq.~(\ref{n-point-first}) beyond the linear order in $\Lambda^4$. This conjecture is reinforced by the fact that Eq.~(\ref{distribution-2}) acquires the correct nontrivial expression in the limit $f \to + \infty$ and $\Lambda \to + \infty$, while keeping $m \equiv \Lambda^2/f$ fixed. Here, $m$ is the mass of $\psi$ at the stable value $\psi = 0$; that is, $m^2 = v''(\psi)\big|_{\psi=0}$.

\begin{figure}[t!]
\includegraphics[scale=0.4]{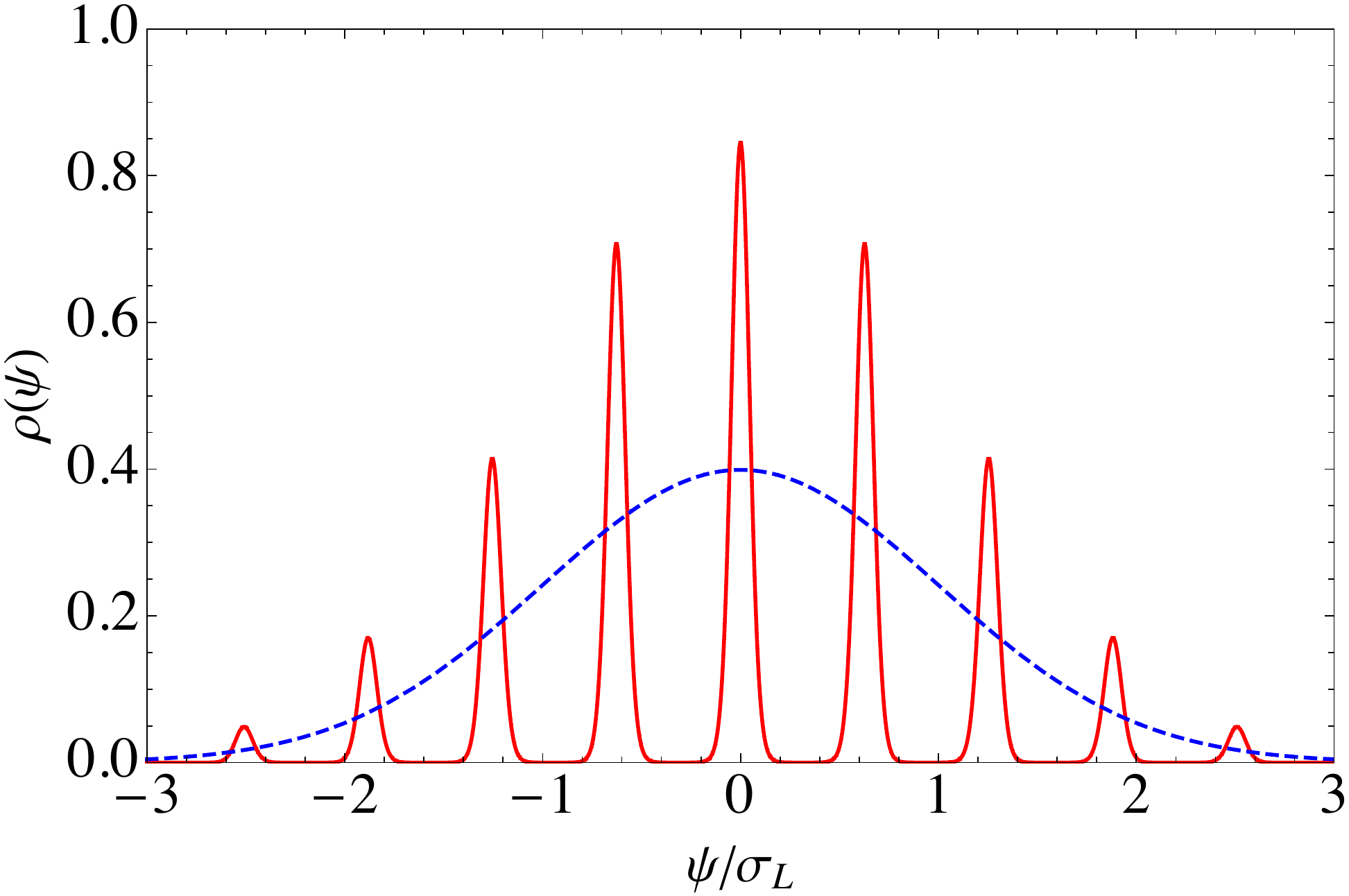}
\caption{The figure shows an example of the PDF of Eq.~(\ref{distribution-2}) for the choice of parameters $f / \sigma_L = 10^{-1}$ and $A^2 / \sigma_L^2 = 10^{-1}$ (red solid curve). For comparison, we have plotted a Gaussian distribution of variance $\sigma_L$ (blue dashed curve). The distributions are not normalized.}
\label{fig:FIG2}
\end{figure}

The PDF of Eq.~(\ref{distribution-2}) is plotted in Fig.~\ref{fig:FIG2} for certain values of $A$ and $f$. The function consists of a multimodal distribution where $\Delta \psi = 2 \pi f$ determines the distance between consecutive peaks. On the other hand, $A$ controls the magnitude with which the probability of measuring a value of $\psi$ laying in the vicinity of a minimum or maximum of $v(\psi)$ is enhanced or suppressed respectively. Notice that the non-Gaussian effects due to the periodicity of the potential are relevant only if $f < \sigma_L$. Given that $\sigma_L$ is of the order of $H$, our result calls for sub-Planckian values of $f$, which in fact are favored in string theory~\cite{Banks:2003sx, Svrcek:2006yi} and quantum gravity~\cite{Conlon:2012tz}. In what follows, we discuss three possible consequences of our result that might be worth exploring.

\subsection{Isocurvature fluctuations after inflation}

It is common lore that scalar fields with large masses during inflation lead to the production of suppressed levels of isocurvature modes in the CMB. Our results show that this is not necessarily the case: If fields are able to tunnel the potential barriers separating their VEV from other minima, then the PDFs describing them at the end of inflation can be highly non-Gaussian, regardless of how massive they are. In the case that we have studied, the mass of the axion about any local minima is given by $m = \Lambda^2/f$, and a large value (as compared to $H$) does not preclude fluctuations from leaking from one minimum to another. Thus, a landscape with a rich structure may have strong quantum effects even with massive extra scalar fields. It would be interesting to study how these effects could affect current studies of inflation with many fields such as those of Refs.~\cite{McAllister:2012am, Marsh:2013qca, Dias:2016slx}, and CMB observables.

\subsection{Role of inflation to determine SM properties}

Our results also reinforce the intriguing possibility that the SM of particle physics could be just one realization among many others, taking place in a confined region of our Universe~\cite{Linde:1993xx}. This would be the case if the field $\psi$ determines the value of couplings of other fields that appear in the SM, such as the Higgs field. The PDF of Eq.~(\ref{distribution-2}) tells us that different patches of the Universe could emerge out of inflation with a long-wavelength value of $\psi$ corresponding to a minimum of $v(\psi)$ different from $\psi=0$, which, by assumption, was the vacuum expectation value of the field. Our results should, in principle, allow one to compute the probability with which these patches emerge.

\subsection{Dark matter}

Another interesting possibility is that $\psi$ corresponds to dark matter (DM)~\cite{Preskill:1982cy, Abbott:1982af, Dine:1982ah}. If this is the case, our result predicts that, for certain values of the parameters, it could be possible that within our observable Universe the DM distribution started with nontrivial initial conditions, resulting from the multimodal PDF~(\ref{distribution-2}). We foresee that, if the initial conditions for DM come from~(\ref{distribution-2}), then there should be certain observational signatures that could serve as a test. It is a pending challenge to deduce them. 

\section{Conclusions}

In summary, we have analyzed the dynamics of the fluctuations of axion spectator fields present during inflation. Our main result consisted in the derivation of a probability distribution function of these fields. This result shows how these spectator fields are able to tunnel between the various vacua of the axionlike potential~\footnote{It would be interesting to understand the relation between our results and other well known tunneling solutions in expanding backgrounds, such as the Hawking-Moss solution~\cite{Hawking:1981fz}.}.  We have limited our analysis to the derivation of this distribution in the long-wavelength limit, omitting for now a more insightful look into the potentially large range of implications that such a result could have. A pending task is the derivation of a distribution incorporating the spatial dependence of the fields. This would allow one, for instance, to study the initial conditions of the inhomogeneous distribution of DM as discussed in the previous section. It should be possible to study our derived PDF with alternative methods, such as the stochastic approach~\cite{Starobinsky:1982ee, Starobinsky:1986fx, Nambu:1987ef} (see, for example,~\cite{Hardwick:2017fjo} for a recent discussion where the case of axionic fields is examined).

\begin{acknowledgments}

We are grateful to Diego Blas, Jinn-Ouk Gong, Jorge Nore\~na and Spyros Sypsas for useful discussions leading to this version of the work. This work is partially supported by the Fondecyt Project No. 1130777. W.R. acknowledges support from the DFI postgraduate scholarship program.

\end{acknowledgments}

\end{document}